\documentclass[aps,prd,draft,showpacs,eqsecnum,twocolomn,tighten]{revtex4}

\begin{document}

\title{Relativistic theory of elastic deformable
astronomical bodies: perturbation equations in rotating spherical
coordinates and junction conditions}

\author{Chongming Xu}
 \email{cmxu@njnu.edu.cn}
\author{Xuejun Wu}
 \email{xjwu@njnu.edu.cn}
\affiliation{Department of physics, Nanjing Normal University,
 Nanjing 210097, China}
\author{Michael Soffel}
 \email{soffel@rcs.urz.tu-dresden.de}
\author{Sergei Klioner}
 \email{klioner@rcs.urz.tu-dresden.de}
\affiliation{Lohrmann
Observatory, Technical University Dresden, D-01062 Dresden,
Germany}

\date{\today}


\begin{abstract}
In this paper, the dynamical equations and junction conditions at
the interface between adjacent layers of different elastic
properties for an elastic deformable astronomical body in the
first post-Newtonian approximation of Einstein theory of gravity
are discussed in both rotating Cartesian coordinates and rotating
spherical coordinates. The unperturbed rotating body (the ground
state) is described as uniformly rotating, stationary and
axisymmetric configuration in an asymptotically flat space-time
manifold. Deviations from the equilibrium configuration are
described by means of a displacement field. In terms of the
formalism of relativistic celestial mechanics developed by Damour,
Soffel $\&$ Xu, and the framework established by Carter $\&$
Quintana the post Newtonian equations of the displacement field
and the symmetric trace-free shear tensor are obtained.
Corresponding post-Newtonian junction conditions  at interfaces
also the outer surface boundary conditions are presented. The PN
junction condition is an extension of Wahr's one which is a
Newtonian junction conditions without rotating.
\end{abstract}


\maketitle


\section{INTRODUCTION}

The theory of elastic deformable bodies is of great importance for
quantitative models for the free and forced motions of
astronomical bodies (specially for the Earth). Historically the
perturbed Newtonian Euler equation for an elastic deformable Earth
was apparently first derived by Jeffreys and Vicente\cite{jeff}.
Some different forms of it have appeared later in the
literature\cite{mori}. Such a local treatment of global
geodynamics has been pursued especially by Wahr\cite{wahr},
Schastok\cite{scha}, and Dehant \& Defraigne\cite{deha} to
describe the nutation of the Earth. Clearly all of these
investigations just mentioned are fully within Newton's theory of
gravity. The theory of elasticity is also of importance for the
interpretation of data resulting from modern observational
techniques such as Very-Long Baseline Interferometry (VLBI), Lunar
and Satellite Laser Ranging (LLR $\&$ LSR), and all other kinds of
observations where the positions of Earth-bound points should be
described with high position. The normal modes (or quasi-normal
modes) of the Earth or other astronomical bodies, such as white
dwarfs or neutron stars is another field of important
applications\cite{hugh,chin,card} as also the  calculation of the
time evolution of the (mass and current) multipole
moments\cite{xu97} of astronomical bodies.

Extending the Newtonian theory of motion of elastic deformable
bodies to include relativistic effects presents a new and improved
basis for further discussions of the such problems. Recently Xu,
Wu and Soffel\cite{xu01} developed such a general relativistic
theory of elastic deformable astronomical bodies on the basis of
the Damour-Soffel-Xu (DSX) formalism as the foundation of modern
general relativistic celestial mechanics at the first
post-Newtonian  approximation of Einstein's theory of gravity
\cite{damo91,damo92,damo93,damo94}. In \cite{xu01} we discussed
the post-Newtonian perturbations of a uniformly rotating,
stationary, and axisymmetric elastic body in a rotating Cartesian
coordinate system. The general perturbations of such a
configuration are treated within the Carter-Quintana
formalism\cite{cart72,cart73}.  A central result was the
post-Newtonian dynamical equation for the displacement field in
Cartesian coordinates representing the post-Newtonian version of
the well-known Jeffreys-Vicente equation\cite{mori}. However, for
practical applications, the common way of dealing with such
perturbations is to go to spherical \cite{lapw} instead of
Cartesian coordinates and then to expand the relevant quantities
in terms of scalar, vector and tensor  spherical harmonics or
so-called generalized spherical harmonics.

In this paper, we follow the route of our previous paper
\cite{xu01} to deduce the post-Newtonian perturbed local evolution
equations  and the perturbed Eulerian equation for the
displacement field of an elastic astronomical body in rotating
spherical coordinates. To this end the Eulerian variation of
Einstein's energy-momentum conservation law is performed. The
Newtonian version of our results (neglecting all $1/c^2$ terms) is
in agreement with standard results from the literature (e.g.,
Ref.\cite{lapw} after correction of a typographical mistake).
Post-Newtonian junction conditions for the transition from one
layer to another  with different elastic properties  that were not
treated in \cite{xu01} are also presented here. Such general
relativistic junction conditions have been discussed in a
different context\cite{lich,robs}. However, the application to a
post-Newtonian displacement field presented here is new.
Corresponding Newtonian junction conditions can e.g., be found in
Wahr\cite{wahr}.

Symbols and notations are taken from the DSX papers
\cite{damo91,damo92,damo93,damo94}: the space-time signature  is
taken as $-+++$, spacetime indices go from 0 to 3 and are denoted
by Greek indices, while spatial indices (1 to 3) are denoted by
Latin indices. We use Einstein's summation convention for both
types of indices, whatever the position of repeated indices. We
shall often abbreviate the order symbol $O(c^{-n})$ simply by
$O(n)$. Local coordinates $X^\alpha=(cT,X^a)$ are chosen in that
reference system that moves with the body under consideration. The
DSX scheme provides a description of the metric tensor
$G_{\alpha\beta}$ in a local system with two metric potentials
$(W, W^a )$.

In Section II, the post-Newtonian perturbation  equations of elastic astronomical
bodies both in Cartesian \cite{xu01} and spherical
coordinates are presented.
Special emphasis is given to the new derivation in spherical
coordinates.
In Section III, we discuss
post-Newtonian junction  and boundary conditions.
In the last Section  some conclusion can be found.


\section{Post-Newtonian perturbation equations of elastic deformable astronomical
bodies}

\subsection{Perturbed equations in Rotating Cartesian coordinates}

The formalism starts by considering some isolated relaxed body
that rotates uniformly with angular velocity ${\bf \Omega}$ about
its symmetry axis with respect to some global non-rotating
coordinate system. Deviations from such an equilibrium
configuration and the action of tidal forces in a gravitational
$N$-body system are then described by means of perturbation
theory. In that case one has to deal with at least three different
coordinate systems: a global coordinate system $x^\mu = (ct,x^i)$
like the Barycentric Celestial Reference System (BCRS) that
extends to infinity and where the dynamics of the $N$-body system
can be formulated, a local 'non-rotating' system $X^\alpha = (cT,
X^a)$ like the Geocentric Celestial Reference System (GCRS) (e.g.,
kinematically non rotating or at least slowly rotating with an
angular velocity of post-Newtonian order with respect to the
global system) and finally some local coordinate system
($\overline{X}^\alpha = (c \overline{T}, \overline{X}^a)$ with $
\overline{T} = T$) whose spatial coordinates co-rotate uniformly
with the equilibrium configuration. The post-Newtonian
perturbation equations of an elastic deformable astronomical body
in rotating Cartesian coordinates  have been presented
recently\cite{xu01}. Here, some of the results are summarized. The
components of  the metric tensor in rotating Cartesian coordinates
read
\begin{eqnarray}
\overline{G} _{00} &=& - \exp \left( - \frac{2 \overline{W}
 + \overline{V}^2}{c^2} \right)+ O(5) \, ,  \\
\overline{G} _{0a} &=& \frac{\overline{V} ^a}{c}
  - \frac{4\overline{W}_a}{c^3} + O(5) \, , \\
\overline{G} _{ab} &=& \delta _{ab} \exp \left(
  \frac{2\overline{W} }{c^2} \right) + O(4) \, ,
\end{eqnarray}
where
\begin{equation}\label{sW}
\overline{W} \equiv W  + \frac{2 W V^2}{c^2}
 - \frac{4 W^b V^b}{c^2}
 + \frac{V^4}{4c^2}
\end{equation}
and
\begin{equation}\label{vW}
\overline{W} _a \equiv R ^{ab} \left( W _b - {1 \over 2} V^b W
   \right) \, .
\end{equation}
A bar on top of some quantity indicates that it refers to the
rotating coordinate system $\overline X^\alpha$, otherwise it will
refer to the 'non-rotating' local coordinate system $X^\alpha$.
 $W$ and $W_a$ are the scalar and vector potential
which describe the metric in local 'non-rotating' coordinates. For
the relaxed ground state of the body these potentials result
entirely from the gravitational action of the body itself. For
more details the reader is referred to Damour et al.\cite{damo91}.
$V^b$ is the rotation velocity of the equilibrium configuration,
$R^{ab}$ is a time-dependent rotation matrix (defined by
(2.15)--(2.16) of \cite{xu01}), $\overline{V}^a = R^{ab}V_b$ and
$\overline{\Omega} ^b = R^{bc}\Omega ^c$. $\Omega ^c$ is the
angular velocity with respect to 'non-rotating' coordinates
($V^a=\epsilon _{abc}\Omega ^b X^c$, $\epsilon_{abc} $ is the
completely antisymmetric Levi-Civita symbol of rank 3 with
$\epsilon_{123} = +1$). The perturbed energy balance equation
after a first time integration reads (see (4.30) and (3.31) of
\cite{xu01})
\begin{eqnarray}\label{drho}
\delta \rho &=& - \nabla \cdot ( \rho {\bf s} ) -
  {1 \over {c^2}} \left( \rho \overline{V} ^a \dot{s}^a
  + (p s ^a ) _{,a} + 2 \overline{W} _{,a} s ^a \rho
  + 3 \rho \delta \overline{W} \right) + O(4) \nonumber \\
 &=&  - \rho _{,a} s^a - \rho ^* \Theta + O(4)
  \, ,
\end{eqnarray}
where $\rho$ is the energy density, $\rho ^* = \rho + p/c^2$ the
chemical potential per unit volume, $p$ is the isotropic pressure
and  $s^a$ are the spatial components of the contravariant
displacement field. The volume dilatation $\Theta$ is given by
\begin{equation}
\Theta = s ^b {}_{,b} + {1 \over {c^2}}
 \left( \overline{V} ^b \dot{s}^b
 + \epsilon _{bcd} \overline{\Omega} ^c s ^d \overline{V} ^b
 + 3 \overline{W} _{,c} s^c
 + 3 \delta \overline{W} \right) + O(4) \, .
\end{equation}
Here, $\delta \overline{W}$ is the Eulerian variation of
$\overline W$ in rotating coordinates. The perturbed Eulerian
equation takes the form (Eq.(4.32) of \cite{xu01})
\begin{eqnarray}\label{euler}
0 &=& \rho ^* \left( 1 + \frac{2 \overline{W}_{\rm G}}{ c^2}
  \right) \left( \ddot{s} _a
  + 2 \epsilon _{abc} \overline{\Omega}^b \dot{s}^c \right)
  + \rho^* \Theta \overline{W}_{{\rm G},a}
  - \rho ^* s^b _{,a} \overline{W}_{{\rm G},b}
  - \rho ^* ( \delta \overline{W}_{\rm G} ) _{,a}
  - \rho ^* s^b \overline{W}_{{\rm G},b a} \nonumber \\
 && \hskip-3mm - ( \kappa \Theta \delta _{a\beta}
  + 2 \mu s ^\beta {}_a ) _{;\beta}
  + \frac{1}{c^2} \left\{ \rho ^* \left[
   \overline{V} ^a (\overline{V} ^b \ddot{s} ^b )
  + \overline{W}_{{\rm G},b} \dot{s}^b \overline{V} ^a
  - 2 \overline{V} ^b \dot{s}^b \overline{W}_{{\rm G},a}
  + ( \delta \overline{W})_{, \overline{T}}
  \overline{V} ^a \right. \right. \nonumber \\
 && \hskip-3mm \left. \left. + 8 \dot{s}^b \overline{W}_{[b,a]}
  - 4 ( \delta\overline{W} _a ) _{, \overline{T}} \right]
  + \kappa \left( \Theta \overline{W}_{{\rm G},a}
  - \dot{\Theta} \overline{V} ^a \right)
  + \left( \kappa ( 4 \overline{W} \Theta
  + \overline{V} ^b \overline{V} ^c s ^b _{,c} ) \right)
  _{,a} \right\} + O(4) \, ,
\end{eqnarray}
where the post-Newtonian geopotential $\overline{W} _{\rm G}$ is
given by
\begin{displaymath}
\overline{W} _{\rm G}
 = \overline{W} + {\overline{V} ^2 \over 2} \, .
\end{displaymath}
The  ``dot" stands for the derivative with respect to the  time
variable $\overline T = T$ and $\delta$ indicates the Eulerian
variation. The elastic moduli $\kappa$ and $\mu$ are the
compression modulus and the shear modulus respectively; $s_{ab}$
is the shear-stress tensor (a complete representation of $s_{ab}$
is given in (4.26) of \cite{xu01}). The  term containing the
shear-stress tensor reads (see also (4.33) of \cite{xu01})
\begin{eqnarray}
(2 \mu s^\beta {}_a ) _{;\beta} &=& ( 2 \mu s_{ba}) _{,b}
 + {1 \over c^2} \left\{ - ( 4\mu \overline{W} s_{ba}
 + 2 \mu s _{ca} \overline{V} ^c \overline{V} ^b ) _{,b}
 + ( 2 \mu s_{ab} )_{,T } \overline{V} ^b
 \right. \nonumber \\
 && \left.
 + 2 \mu ( 2 \overline{W}_{,c} s _{ac} + \epsilon _{acb} \Omega ^c
 s_{bd}\overline{V} ^d ) \right\} \, .
\end{eqnarray}
Equation (\ref{euler}) is the post-Newtonian Euler equation for the
displacement field $\bf s$, sometimes called the post-Newtonian
Jeffreys-Vicente equation.


\subsection{Perturbation equations in rotating spherical
coordinates}

\subsubsection{Unperturbed and perturbed projection tensor in rotating
spherical coordinates}

The importance to formulate the perturbation equations in rotating
spherical coordinates was already stressed in the Introduction.
Note that both the Eulerian and the Lagrangian variation of a
tensor do not necessarily preserve the tensor character, whereas
the difference between them, the Lie derivative (${\cal L} _\xi =
\Delta - \delta $), does. This implies that one cannot simply
transform the variational equations from Cartesian to spherical
coordinates. For that reason all derivations of the perturbed
energy equation and the post-Newtonian Jeffreys-Vicente equation
have to be repeated in spherical coordinates. These rotating
spherical coordinates $(r,\theta,\phi)$ are defined by $(\overline
X = r \sin\theta \cos\phi, \overline Y = r \sin\theta \sin\phi,
\overline Z = r \cos\theta)$.

The metric tensor in such rotating spherical coordinates
($\widetilde{X}^\mu = (c\widetilde{T},r,\theta,\phi)$ with
$\widetilde{T} = T$) takes  the form
\begin{eqnarray}
\label{mett}
\widetilde{G} _{00} &=& - \exp\left( - \frac{2\widetilde{W} +
  \widetilde{V}^2}{c^2} \right) + O(6) \, , \\
\widetilde{G} _{0a} &=& \widetilde{G} _{a0} =  D_{ac} \left(
  \frac{\widetilde{V} ^c}{c}
 - \frac{4\widetilde{W} ^c}{c^3} \right) + O(5) \, , \label{smet}  \\
\widetilde{G} _{ab} &=& D_{ab} \exp \left( \frac{2\widetilde{W}
  }{c^2} \right) + O(4) \, ,
\end{eqnarray}
where $\widetilde{V} ^a= \left( \partial \widetilde{X }^a /
\partial \overline{X} ^b \right) \overline{V} ^b $,
$\widetilde{W} ^a = \left( \partial \widetilde{X} ^a /
\partial \overline{X} ^b \right)\overline{W} ^b  $,
$\widetilde{W} = \overline{W}$, $\widetilde{V}^2 = \overline{V}^2$
and
\begin{equation}
D_{ab} \equiv
 \left( \begin{array}{ccc}
 1 & 0 & 0 \\
 0 & r^2 & 0  \\
 0 & 0 & r^2 \sin ^2 \theta
 \end{array}  \right) \, .
\end{equation}
The corresponding inverse matrix $D^{ab}$ satisfies $D_{ab} D^{bc}
= \delta^c_a$. For the 3-dimension quantities $\widetilde{V} ^a$
and $\widetilde{W} ^a$ we define $ \widetilde{V} _a = D _{ab}
\widetilde{V} ^b $, $ \widetilde{W} _a = D _{ab} \widetilde{W}^b$.

\medskip\noindent
In the following we will choose ${\bf \Omega} = \Omega \, {\bf
e}_z$, with $\Omega $ being constant, so that
\begin{equation}
\widetilde{V}^r = \widetilde{V}_r = \widetilde{V}^\theta =
\widetilde{V}_\theta = 0 \, , \quad \widetilde{V} ^\phi = \Omega
\, , \quad \widetilde{V} _\phi = \Omega \, r^2 \sin ^2\theta \,
\end{equation}
and $ \widetilde{V} ^2 = \Omega ^2 r ^2 \sin ^2\theta$. From these
quantities the Christoffel symbols and the orthogonal projection
tensor (projecting into the 3-space of an observer that moves with
the corresponding material element) can be derived. In Carter and
Quintana's formalism\cite{cart72} a body is described by means of
a bundle of time-like world-lines in a four-dimensional space-time
manifold. A mapping into a three-dimensional manifold that is
composed of the various material elements of the body identifies
the various ``particles" of the body. The projection of any tensor
onto the local rest frame of matter is achieved with the
orthogonal projection tensor
\begin{equation}
\gamma _{\mu\nu} =
G _{\mu\nu} + {U_\mu U_\nu \over c^2} \, .
\end{equation}
This tensor acts as a positive metric
tensor on the tangent subspace orthogonal to the flow vectors. The
corresponding inverse metric tensor is given simply by its
equivalent contravariant form, since it satisfies $\gamma
^{\mu\nu} \gamma _{\nu\sigma} = \gamma ^\mu _\sigma $. $U ^\mu$ is
the four-velocity of some material element as tangent vector to its world
line and $\gamma _{\mu\nu} U ^\nu = 0 $.

\bigskip
In General Relativity, the unperturbed and perturbed states of a
body are considered as two configurations in separate
four-dimensional space-time manifolds. Usually one starts with
canonical coordinates $x^\mu$ in both manifolds and $x ^\mu
\rightarrow x ^\mu + \Delta x ^\mu$ maps the coordinates of a
material element in the reference state onto the coordinates of
the same element in the perturbed state, where $\Delta x ^\mu $ is
the position coordinate displacement in 4-dimension space-time.
The quantities $ \xi ^\mu \equiv \Delta x ^\mu $ are called the
four-dimensional displacement field. The Lagrangian variation is
the variation of the field in terms of a coordinate system which
is itself dragged along by the displacement $\Delta x ^\mu$; it is
denoted by the symbol $\Delta $. The Eulerian variation denoted by
$\delta$ is the variation taken at a fixed point in 4-dimension
space-time. The relation between these two kinds of variations is
given by $ \delta = \Delta - {\mathcal{L}} _\xi $,  where
${\mathcal{L}} _\xi $ stands for the Lie derivative along the
displacement field $\xi ^\mu$. The displacement field, for obvious
reasons, will be defined in rotating coordinates. It is taken as
$\xi ^\mu = (0, \xi ^a)$ (see \cite{xu01}).

\medskip
The Euler variation of $\widetilde{G} _{\mu\nu}$ (denoted by
$h_{\mu\nu}$) results from (\ref{mett}):
\begin{eqnarray}
h _{00} &=& \delta \widetilde{G}_{00}
  = \frac {2 \delta{\widetilde{W}}}{c^2} \left( 1
- \frac{2\widetilde{W} + \widetilde{V} ^2}{c^2} \right)
   + O(6) \, , \\
h _{0a} &=& \delta \widetilde{G}_{0a}
  = - D _{ac} \left( \frac{4\delta{\widetilde{W}} _c}{c^3} \right) + O(5) \, ,
  \\
h _{ab} &=& \delta \widetilde{G}_{ab}
  = D_{ab} \left( \frac{2\delta{\widetilde{W}}}{c^2} \right) + O(4) \, .
\end{eqnarray}
Other important quantities are
the Eulerian variations of the 4-velocity and the projection
tensor. They  take the form
\begin{eqnarray}
\delta U ^0 &=& {1 \over c }  \left(
  \delta \widetilde{W} + {\widetilde{V}_a }{\xi^a }_{,T} \right)
  + O(3) \, ,  \\
\delta U ^a &=& \xi ^a{}_{,T} \left( 1 +
  \frac{\widetilde{W} _G}{c^2} \right) + O(4) \, , \\
\delta U _0 &=& {1 \over c }
  \delta \widetilde{W}  + O(3) \, ,  \\
\delta U _a &=& D_{ab} \left\{ \xi ^b {}_{,T}
 \left( 1 + \frac{\widetilde{W} _G}{c^2} \right)
 + {1 \over {c^2}} \left[ 2 \widetilde{W} \xi ^b {}_{,T}
 - 4 \delta \widetilde{W} ^b + \widetilde{V} ^b (\delta \widetilde{W}
 + {\widetilde{V}_a}  {\xi^a} _{,T}) \right] \right\} + O(4)
\end{eqnarray}
and
\begin{eqnarray}
\delta \gamma ^0{}_0 &=& - {1 \over c^2} {\widetilde{V}_a }
 \xi^a _{,T}+ O(4)  \, , \\
\delta \gamma ^0{}_b &=& {1 \over c} D_{bc} \xi ^c{}_{,T}
 + O(3) \, , \\
\delta \gamma ^a{}_0 &=& - {1 \over c} \xi ^a {}_{,T} + O(3) \, , \\
\delta \gamma ^a{}_b &=&  {1 \over {c^2}} \xi ^a{}_{,T}
 D_{bc} \widetilde{V} ^c + O(4) \, .
\end{eqnarray}


\subsubsection{Lagrangian strain tensor and shear tensor}

For a perfect elastic body, the energy-momentum tensor of the
relaxed state is of the form
\begin{equation}
T _{\alpha\beta} = \rho U _\alpha U _\beta
 + p \gamma _{\alpha\beta} \, ,
\end{equation}
where $\rho $ is the rest energy density and $p$ the isotropic
pressure in the reference state.
The perturbed configuration changes the energy-momentum
distribution and geometrical shape with time and might experience
tidal forces from other astronomical objects.
The Eulerian variation of the energy-momentum tensor can be
expressed as
\begin{equation}
\delta T _{\alpha\beta} = U _\alpha U _\beta \delta \rho
 + \rho \delta (U _\alpha U _\beta)
 + \gamma _{\alpha \beta} \delta p + p \delta \gamma _{\alpha\beta}
 - 2 \mu s _{\alpha\beta} \, ,
\end{equation}
where $s _{\alpha\beta}$ is the shear tensor and  $\mu$ is the
shear modulus. The symmetric trace-free shear tensor is defined as
\begin{equation}
s _{\mu \nu} \equiv e _{\mu\nu } - {1 \over 3} \Theta \gamma
  _{\mu\nu} \, ,
\end{equation}
where the volume dilatation is given by $\Theta = e ^\mu {}_\mu $.

\noindent
The Lagrangian strain tensor is defined by
\begin{equation}
e _{\mu\nu} = {1 \over 2} \Delta \gamma _{\mu\nu}
  = {1 \over 2} ( \gamma _{\mu\nu} - \gamma _{\mu\nu} ^* )
 \, ,
\end{equation}
where $\gamma _{\mu\nu} ^*$ is the unstrained value. Restricting
ourselves  to linear perturbations,  the strain tensor is given by
\cite{cart72}
\begin{equation}
\label{emunu}
e _{\mu\nu } = {1 \over 2}
 \gamma _\mu ^\alpha \gamma _\nu ^\beta \left( h_{\alpha\beta }
  + 2 \xi _{(\alpha; \beta)} \right) \, .
\end{equation}

In the following we choose the displacement field in rotating
spherical coordinates as $ \xi ^\beta = ( 0, \xi ^r, \xi ^\theta
,\xi ^\phi)$ and we use the linear displacement
\begin{equation}
{\bf D} \equiv D^r {\bf e}_r + D^\theta {\bf e}_\theta
  + D^\phi {\bf e}_\phi
\end{equation}
with
\begin{equation}\label{linD}
D^r = \xi ^r , \qquad D^\theta = r \xi ^\theta , \qquad  D^\phi =
r \sin\theta \xi ^ \phi \, .
\end{equation}

\noindent
The explicit calculation of the Lagrangian strain tensor
$e_{\mu\nu}$ and the shear tensor $s_{\mu\nu}$ is straightforward
but cumbersome. Results for
$e_{\mu\nu}$ and $s_{\mu\nu}$ to post-Newtonian accuracy
are given in the Appendix.


\subsubsection{The post-Newtonian energy and Eulerian equations}

The Eulerian variation of the pressure $\delta p$ can be derived
similarly as in our previous paper dealing with  Cartesian
coordinates \cite{xu01}. In spherical coordinates it takes the
form
\begin{equation}
\delta p = -\rho ^* {\bf D} \cdot \nabla \widetilde{W} _G
 - \kappa \Theta + \frac{\kappa}{c^2} \left( 4 \Theta \widetilde{W}
 + \Omega ^2 r \sin\theta ( D^\phi _{,\phi}
 + \sin\theta D^r \right.
\left. + \cos\theta D ^\theta )
  \right) + O(4) \, .
\end{equation}

Our main results concern the perturbed local evolution equations
\begin{equation}
\delta \left( \widetilde{T} ^\nu {}_{\mu ;\nu} \right) =0 \, .
\end{equation}

\noindent
For $\mu = 0$ one derives the perturbed energy equation
in the form
\begin{eqnarray}
\delta \rho &=& - {\bf D} \cdot  \nabla \rho
 - \rho \Theta - {1 \over c^2} p \Theta \nonumber \\
 &=& - \left( D^r \rho _{,r}
 + {1 \over r} D^\theta \rho _{,\theta}
 + \frac{1}{r\sin\theta} D^\phi \rho _{,\phi} \right)
 - \rho ^* \Theta  + O(4) \, . \label{sdrho}
\end{eqnarray}
The perturbed Euler equations  correspond to the case $\mu =
a$. As an intermediate result  by using the expression  for
$\nabla \cdot {\bf D}$ from the Appendix we get
\begin{eqnarray*}
\nabla^2 {\bf D} &=& {\bf e}_r \left[ \nabla^2 D^r - {2\over
 r^2}D^r
 - \frac{2}{r^2 \sin\theta } (\sin\theta D^\theta ) _{,\theta}
 - \frac{2}{r^2 \sin\theta } ( D^\phi ) _{,\phi} \right]
  \\
 && \hskip-4mm + \;  {\bf e}_\theta  \left[ \nabla ^2 D^\theta
 - \frac{D^\theta}{r^2 \sin^2\theta}
 + \frac{2}{r^2 } ( D^r ) _{,\theta}
 - \frac{2 \cos\theta}{r^2 \sin^2\theta }
 ( D^\phi ) _{,\phi} \right]  \\
 && \hskip-4mm + \; {\bf e}_\phi \left[ \nabla ^2 D^\phi
 - \frac{D^\phi}{r^2 \sin^2\theta}
 + \frac{2}{r^2 \sin\theta } ( D^r ) _{,\phi}
 + \frac{2 \cos\theta}{r^2 \sin^2\theta }
 ( D^\theta ) _{,\phi} \right] \, ,
\end{eqnarray*}
where ${\bf e}_r, \, {\bf e}_\theta$ and ${\bf e}_\phi $ are unit
vectors in the $r,~\theta$ and $\phi$ direction respectively.
\begin{eqnarray*}
(\nabla \cdot {\bf D} ) _{,r} &=&
 D^r_{,rr} - {1 \over r^2} D^\theta _{,\theta}
 + {1 \over r} D^\theta _{,\theta r}
 + \frac{1}{r \sin\theta} D^\phi _{,\phi r}
 - \frac{1}{r^2 \sin\theta} D^\phi _{,\phi }  \\
 && \hskip-3mm +  {2 \over r} D^r _{,r} - {2 \over r^2} D^r
 + {1 \over r} \cot\theta D^\theta _{,r}
 - {1 \over r^2} \cot\theta D^\theta \, , \\
(\nabla \cdot {\bf D} ) _{,\theta} &=&
 D^r_{,r\theta} + {1 \over r} D^\theta _{,\theta\theta}
 + \frac{1}{r \sin\theta} D^\phi _{,\phi\theta}
 - \frac{\cos\theta}{r \sin^2\theta} D^\phi _{,\phi }
 + {2 \over r} D^r _{,\theta}  \nonumber \\
 && \hskip-3mm + {1 \over r} \cot\theta D^\theta _{,\theta}
 - \frac{1}{r \sin^2\theta} D^\theta \, , \\
(\nabla \cdot {\bf D} ) _{,\phi} &=&
 D^r_{,r\phi}
 + {1 \over r} D^\theta _{,\theta\phi}
 + \frac{1}{r \sin\theta} D^\phi _{,\phi\phi}
 + {2 \over r} D^r _{,\phi}
 + {1 \over r} \cot\theta D^\theta _{,\phi} \, .
\end{eqnarray*}
Tedious calculations then lead to the explicit equations for $\mu
= a= r, \, \theta, \, \phi $ respectively.

\bigskip
The $\mu = r$ equation reads
\begin{eqnarray}
0 &=& \rho ^* \left( 1 + \frac{2 \widetilde{W} _G}{c^2} \right)
 \left( D^r_{,T T} - 2 \Omega \sin\theta D^\phi _{,T} \right)
 + \rho ^* \Theta \widetilde{W} _{G,r}
 - \rho ^* (\delta{\widetilde{W}} _G ) _{,r} \nonumber \\
 && \hskip-3mm - \rho ^* ( {\bf D} \cdot \nabla  \widetilde{W} _G ) _{,r}
  - (\kappa \Theta )_{,r}
 - 2(\mu s^\beta {}_r) _{;\beta }
 + {1 \over c^2}\left\{ \rho ^* \left[ 2\widetilde{W} D^r_{,TT}
 + \frac{8}{r}D^\theta _{,T}\widetilde{W} _{[\theta ,r]}
 \right. \right. \nonumber \\
 && \hskip-3mm \left. \left.
 + \frac{8}{r\sin\theta}D^\phi _{,T}\widetilde{W} _{[\phi ,r]}
 - 2\Omega r\sin\theta \widetilde{W} _{G,r}D^\phi _{,T}
 - 4 (\delta{\widetilde{W}} ^r ) _{,T } \right]
 + \kappa \Theta \widetilde{W} _{G,r}
 \right. \nonumber \\
 && \hskip-3mm \left. + 4 (\kappa \Theta \widetilde{W} )_{,r}
  + \left[
 \kappa \Omega ^2 \left( s_{\phi\phi} + {1 \over 3}
 \Theta r^2 \sin^2\theta \right) \right]_{,r}
  \right\} + O(4) \, , \label{sEuler1}
\end{eqnarray}
where
\begin{eqnarray}
2\left( \mu s^\beta {}_r \right) _{;\beta} &=&
 2 \mu _{,\beta } s^\beta {}_r
 + 2 \mu s^\beta {} _{r;\beta} \nonumber \\
 &=& \mu \left( {1 \over 3} ( \nabla \cdot {\bf D} )_{,r}
 + (\nabla ^2 {\bf D} ) _r \right)
 + 2 \mu _{,r} \left( D^r_{,r}
 - {1 \over 3} \nabla \cdot {\bf D}  \right)
 \nonumber \\
 && \hskip-3mm + \frac{\mu _{,\theta}}{r ^2}\left( D^r_{,\theta}
 + r D^\theta _{,r} - D^\theta \right)
 + \frac{\mu _{,\phi}}{r\sin\theta}
 \left( \frac{D^r_{,\phi}}{r\sin\theta} + D^\phi_{,r}
 - \frac{D^\phi}{r} \right) \nonumber \\
 && \hskip-3mm + {1 \over c^2} \left\{ 2\mu \left[
 2 s_{rr}\widetilde{W} _{,r}
 + \frac{2 s_{r\theta}}{r^2} \widetilde{W} _{,\theta}
 + \frac{2 s_{r\phi}}{r ^2 \sin^2\theta} \widetilde{W} _{,\phi}
 \right. \right. \nonumber \\
 && \hskip-3mm \left. \left. + 2\Omega s_{r\phi,T} - \frac{5}{6}
  \Omega \sin\theta ( r D^\phi _{,T} )_{,r}
 - {1 \over 3}\Omega ^2 r \sin\theta \cos\theta D ^\theta _{,r}
 - \frac{\Omega ^2}{2}D ^r _{,\phi\phi} \right. \right. \nonumber \\
 && \hskip-3mm \left. \left.  - {7 \over 3} \Omega ^2\sin^2\theta
 ( D^r + D^\theta \cot\theta )
 - \Omega ^2 \sin\theta D^\phi_{,\phi} \right] \right. \nonumber \\
 && \hskip-3mm \left. - \frac{2}{3} \mu_{,r}\Omega r \sin\theta  \left( D^\phi _{,T}
 + D^r \Omega \sin\theta + D^\theta \Omega \cos\theta \right)
 + \mu _{,\phi} \Omega \left( D^r_{,T}
 - \Omega D^r_{,\phi} \right)  \right\}
 + O(4)\, . \nonumber \\
 &&  \label{sEuler2}
\end{eqnarray}
Two indices enclosed in  parentheses imply symmetrization as in
$\widetilde{W} _{(a,b)} = ( \widetilde{W} _{a,b} + \widetilde{W}
_{b,a}) /2 $, and two indices enclosed in  square brackets imply
anti-symmetrization as in $\widetilde{W} _{[a,b]} = (
\widetilde{W} _{a,b} - \widetilde{W} _{b,a}) /2 $.

\bigskip
The $\mu = \theta$ equation takes the form
\begin{eqnarray}
\label{restheta} 0 &=& \rho ^* r  \left( 1
 + \frac{2 \widetilde{W}_G}{c^2} \right)
 \left( D^\theta _{,TTt} - 2 \Omega \cos\theta D^\phi _{,T} \right)
 + \rho ^* \Theta \widetilde{W} _{G,\theta}
 - \rho ^* (\delta{\widetilde{W}} _G )_{,\theta} \nonumber \\
 && \hskip-3mm - \rho ^* ( {\bf D} \cdot\nabla \widetilde{W} _G ) _{,\theta}
  - (\kappa \Theta )_{,\theta}
 - 2(\mu s^\beta _\theta) _{;\beta }
 + {1 \over c^2}\left\{ \rho ^* \left[
 2r\widetilde{W} D^\theta _{,T T} + 8 D^r_{,T}\widetilde{W} _{[r, \theta ]}
 \right. \right. \nonumber \\
 && \hskip-3mm \left. \left.
 + \frac{8}{r\sin\theta}D^\phi_{,T}\widetilde{W} _{[\phi ,\theta]}
 - 2\Omega r\sin\theta \widetilde{W} _{G,\theta}D^\phi _{,T}
 - 4 (\delta{\widetilde{W}} _\theta ) _{,T } \right]
 + \kappa \Theta \widetilde{W} _{G,\theta}
 \right. \nonumber \\
 && \hskip-3mm \left. + 4 (\kappa \Theta \widetilde{W} )_{,\theta}
  + \left[ \kappa \Omega ^2 \left( s_{\phi\phi} + {1 \over 3}
 \Theta r^2 \sin^2\theta \right) \right]_{,\theta}
 \right\} + O(4) \, , \label{sEuler3}
\end{eqnarray}
where
\begin{eqnarray}
2\left( \mu s^\beta {}_\theta \right) _{;\beta} &=&
 2 \mu _{,\beta } s^\beta {}_\theta
 + 2 \mu s^\beta {} _{\theta;\beta} \nonumber \\
 &=& \mu \left( {1 \over 3} ( \nabla \cdot {\bf D} )_{,\theta}
 + r (\nabla ^2 {\bf D} ) _\theta \right)
 + \mu_{,r} \left( D^r _{,\theta}
 + r D ^\theta _{,r} - D^\theta \right) \nonumber \\
 && \hskip-3mm + 2 \mu _{,\theta} \left( \frac{D^\theta _{,\theta}}{r}
 + \frac{D^r}{r} -{1 \over 3} \nabla \cdot {\bf D} \right)
 + \frac{\mu _{,\phi}}{r\sin\theta}
 \left( \frac{D^\theta _{,\phi}}{\sin\theta}
 + D^\phi _{,\theta} - D^\phi \cot\theta \right) \nonumber \\
 && \hskip-3mm + {1 \over c^2} \left\{ 2\mu \left[
 2 s_{\theta r}\widetilde{W} _{,r}
 + \frac{2 s_{\theta\theta}}{r^2} \widetilde{W} _{,\theta}
 + \frac{2 s_{\theta\phi}}{r ^2 \sin^2\theta} \widetilde{W} _{,\phi}
 + 2\Omega s_{\theta\phi ,T} - \frac{5}{6}
  \Omega r ( \sin\theta D^\phi _{,T} )_{,\theta}
 \right. \right. \nonumber \\
 && \hskip-3mm \left. \left.
 - {1 \over 2}\Omega ^2 r D^\theta _{,\phi\phi}
 + {1 \over 3} \Omega ^2 r^2 \sin\theta \cos\theta D^r_{,r}
 - {1 \over 3} \Omega ^2 r \sin^2\theta D ^r _{,\theta}
 - {2 \over 3} \Omega ^2 r \cos\theta D ^\phi_{,\phi}
 \right. \right. \nonumber \\
 && \hskip-3mm \left. \left.
 - 2 \Omega ^2 r \sin\theta \cos\theta D^r
 + {1 \over 3} \Omega ^2 r^2 \sin^2\theta D^\theta
 - 2 \Omega ^2 r \cos^2\theta D^\theta \right] \right.
 \nonumber \\
 && \hskip-3mm \left. - \frac{2}{3} \mu _{,\theta} \Omega r \sin\theta
  \left( D^\phi _{,T} + \Omega ( D^r \sin\theta
  + D^\theta \cos\theta ) \right)
 \right. \nonumber \\
 && \hskip-3mm \left. + \mu _{,\phi} \Omega r \left( D^\theta _{,T}
 - \Omega D^\theta _{,\phi} \right) \right\} + O(4) \, .
 \label{sEuler4}
\end{eqnarray}

\bigskip
Finally, the $\mu = \phi$ equation reads explicitly
\begin{eqnarray}
\label{resphi} 0 &=& \rho ^* r \sin\theta \left( 1 + \frac{2
\widetilde{W} _G}{c^2} \right)
 \left( D^\phi _{,T T} + 2 \Omega \sin\theta (D^r _{,T}
 + \cot\theta D^\theta _{,T}) \right) \nonumber \\
 && \hskip-3mm + \rho ^* \Theta \widetilde{W} _{G,\phi}
 - \rho ^* (\delta{\widetilde{W}} _G ) _{,\phi}
 - \rho ^* ( {\bf D} \cdot \nabla  \widetilde{W} _G ) _{,\phi}
 - (\kappa \Theta )_{,\phi}
 - 2(\mu s^\beta {}_\phi) _{;\beta } \nonumber \\
 && \hskip-3mm + {1 \over c^2}\left\{ \rho ^* \left[
 2r\sin\theta \widetilde{W} _G D^\phi _{,T T}
 + \Omega r^2 \sin^2\theta ({\bf D}_{,T } \cdot \nabla \widetilde{W} _G )
 - 2\Omega r \sin\theta \widetilde{W} _{G,\phi} D ^\phi _{,T}
 \right. \right. \nonumber \\
 && \hskip-3mm \left. \left. + 8 D^r_{,T}\widetilde{W} _{[r, \phi ]}
 + \frac{8}{r}D^\theta_{,T}\widetilde{W} _{[\theta , \phi]}
 + \Omega r ^2 \sin^2\theta (\delta{\widetilde{W}} )_{,T}
 - 4 (\delta{\widetilde{W}} _\phi ) _{,T } \right] \right. \nonumber \\
 && \hskip-3mm \left. + \kappa ( \Theta \widetilde{W} _{G,\phi}
 - \Theta _{,T} \Omega r^2 \sin^2\theta )
 + 4 (\kappa \Theta \widetilde{W} )_{,\phi} + \left[
 \kappa \Omega ^2 \left( s_{\phi\phi} + {1 \over 3}
 \Theta r^2 \sin^2\theta \right) \right]_{,\phi}
  \right\} + O(4) \, , \nonumber \\
 && \label{sEuler5}
\end{eqnarray}
where
\begin{eqnarray}
2\left( \mu s^\beta {}_\phi \right) _{;\beta} &=&
 2 \mu _{,\beta } s^\beta {}_\phi
 + 2 \mu s^\beta {} _{\phi;\beta} \nonumber \\
 &=& \mu \left( {1 \over 3} ( \nabla \cdot {\bf D} )_{,\phi}
 + r\sin\theta (\nabla ^2 {\bf D} ) _\phi \right)
 + \mu_{,r} r \sin\theta \left(
 \frac{D^r _{,\phi}}{r \sin\theta}
 + D ^\phi _{,r} - \frac{D^\phi }{r} \right) \nonumber \\
 && \hskip-3mm +  \mu _{,\theta} \sin\theta \left(
 \frac{D^\phi_{,\theta}}{r}
 + \frac{D^\theta _{,\theta}}{r\sin\theta}
 - \frac{D^\phi \cot\theta }{r} \right) \nonumber \\
 && \hskip-3mm + 2 \mu _{,\phi} \left(
 \frac{D^\phi _{,\phi}}{r\sin\theta}
 + \frac{D^r}{r} + \frac{D^\theta \cot\theta }{r}
 -{1 \over 3} \nabla \cdot {\bf D} \right) \nonumber \\
 && \hskip-3mm + {1 \over c^2} \left\{ 2 \mu \left[
 2 s_{\phi r} \widetilde{W} _{,r}
 + \frac{2 s_{\phi\theta}}{r^2} \widetilde{W} _{,\theta}
 + \frac{2 s _{\phi\phi}}{r ^2 \sin^2\theta} \widetilde{W} _{,\phi}
 + {1 \over 6} \Omega r^2 \sin^2\theta (\nabla \cdot {\bf D})_{,T}
 \right. \right. \nonumber \\
 && \hskip-3mm \left. \left.
 + {1 \over 2} \Omega ^2 r^3\sin^3\theta (\nabla ^2 {\bf D})_\phi
 +{7 \over 6} \Omega r \sin\theta D^\phi _{,\phi t} \right.
 \right. \nonumber \\
 && \hskip-3mm \left. \left. + 2\Omega r \sin\theta ( D^r_{,T }\sin\theta
 + D^\theta _{,T } \cos\theta )
 - {1 \over 3} \Omega^2 r\sin\theta \left( D^r_{,\phi}\sin\theta
 + D^\theta _{,\phi} \cos\theta \right) \right. \right. \nonumber \\
 && \hskip-3mm \left. \left. + {1 \over 2} \Omega^2 r \sin\theta \left(
 r \sin^2\theta D^\phi _{,r} - D^\phi
 + 2 \cos\theta D^\phi _{,\theta} - D^\phi _{,\phi\phi}
 \right) \right] \right. \nonumber \\
 && \hskip-3mm \left. + \mu_{,r}\Omega r^2 \sin^2\theta \left[
 \Omega r \sin\theta \left( \frac{D^r _{,\phi}}{r \sin\theta}
 + D ^\phi _{,r} - \frac{D^\phi }{r} \right)
 + D^r_{,T } - \Omega D^r_{,\phi}  \right] \right. \nonumber \\
 && \hskip-3mm \left. +  \mu _{,\theta} \Omega r \sin^2\theta \left[
 \Omega r \sin\theta \left( \frac{D^\phi_{,\theta}}{r}
 + \frac{D^\theta _{,\theta}}{r\sin\theta}
 - \frac{D^\phi \cot\theta }{r} \right)
 + D^\theta _{,T } - \Omega D^\theta _{,\phi}
 \right] \right. \nonumber \\
 && \hskip-3mm \left. + \frac{4}{3}\mu _{,\phi} \Omega r \sin\theta
 \left( D^\phi _{,T } + \Omega ( D^r\sin\theta
 + D^\theta \cos\theta ) \right) \right\}
 + O(4) \, . \label{sEuler6}
\end{eqnarray}

\bigskip
Eqs.(\ref{sEuler1}), (\ref{sEuler3}), (\ref{sEuler5}) together
with (\ref{sdrho}) are the desired dynamical equations for the
displacement field. They are valid up to terms of order $1/c^4$
and second order in the displacements field itself. The Newtonian
limit of our results (neglecting all $ 1 / c^2$ terms) agrees with
standard results from textbooks (e.g., the ones from \cite{lapw}
after correction of a  typographical mistake).


\section{Post-Newtonian junction conditions}

In most cases such post-Newtonian dynamical equations of elastic
deformable bodies will  be applied to astronomical bodies composed
of different layers. If the body has several layers as e.g., the
Earth that shows a solid inner core, a fluid outer core, a mantle
and a thin crust, we have to consider corresponding junction
conditions at the interface of two adjacent layers. First we
consider such junction conditions in Cartesian coordinates. For
practical applications  they are then also formulated in spherical
coordinates where they can be compared with well-known Newtonian
results \cite{wahr}. Junctions conditions are formulated in
rotating coordinates and to have a well defined stationary and
axisymmetric ground state it is assumed that all layers rotate
with the same angular velocity ${\bf \Omega}$, i.e. there is no
relative motion between two layers in the ground state. The
behavior of individual physical quantities and their corresponding
perturbed quantities on the interface will be studied at first.

The gravitational potentials $W$ and $W^a$ are physical quantities
in the non-rotating ground state. For a isolated body, $W$ and
$W^a$ can be obtained as a solution of Eqs.(2.4), (2.5) in
\cite{xu01}. They are inhomogeneous D'Alembert's and Poisson
differential equation respectively. Although the sources $\Sigma$
and $\Sigma^a$ may be discontinuous across any interface, $W$ and
$W^a$ (solutions of the equations) are continue on the interface.
Since we do not consider the shock wave, the surface mass-density
and surface current mass-density do not exist anywhere. Therefore
derivatives of $W$ and $W^a$ have to be continue on the interface
also. ${\overline W}$ and ${\overline W}^a$ in rotating
coordinates differ from ($W, \, W^a$) simply by a nonsingular
coordinate transformation (Eq.(\ref{sW}) and (\ref{vW})). So
$\overline W$, ${\overline W}^a$ and their derivative continuous
on any interface as well.

To discuss the continuity of $\delta {\overline W}$ and $\delta
{\overline W}^a$ we have to consider $\delta W $ and $\delta W^a$
in non-rotating coordinates before hand, since only in
non-rotating coordinates $\delta W$ and $\delta W^a$ can be
obtained from perturbed field equations (Eq.(4.17) and (4.18) of
Ref.\cite{xu01})
\begin{eqnarray}
\left( \nabla ^2 - {1 \over c^2} \frac{\partial ^2}{\partial T^2 }
\right) \delta W
 &=& - 4\pi G \delta \Sigma + O(4) \label{nab1} \\
\nabla ^2 \delta W ^a &=& - 4\pi G \delta \Sigma ^a + O(2)
 \label{nab2}
\end{eqnarray}
where
\begin{eqnarray}
\delta  \Sigma &=& - \rho^*_{,a} s^a - \rho ^* \Theta
  + \frac{1}{c^2} \left[ 2 \rho^* ( \delta \overline{W}
  + 2\overline{V}^a \dot{s}^a)
  - 2 (\rho^*_{,a}s^a  +\rho ^* \Theta )(\overline{W}
  + \overline{V}^2)  \right. \nonumber \\
 && \left.  -2\rho ^* \overline{W}_{{\rm G},b} s^b
 - 3\kappa \Theta \right] + O(4) \, , \label{dSig} \\
\delta \Sigma ^a &=& R^{ba}\left[ \rho^* \dot{s}^b
 - \overline{V}^b (\rho^*_{,c} s^c + \rho^*\Theta ) \right]
 + O(2) \, . \label{dSiga}
\end{eqnarray}
$\delta W$ and $\delta W^a$ are solutions of Eqs.(\ref{nab1}),
(\ref{nab2}). They are continuous as the same discussion on $W$
and $W^a$, but their derivatives are in different cases. The
Eulerian variation of the surface mass-density and Eulerian
variation of the surface current mass-density do exist on the
interface because both $\delta \Sigma$ and $\delta \Sigma^a$ are
dependent on the spatial coordinate derivatives of $\rho$ (to see
Eq.(\ref{dSig}), (\ref{dSiga})). Then on interface $\delta\Sigma$
and $\delta\Sigma^a$ are divergence. Therefore $(\delta W)_{,a}$
and $(\delta W^a)_{,b}$ are finite on the interface but not
necessary to be continuous across the interface. Through the
coordinate transformation and neglecting all of higher order
terms, we get the representation of $\delta {\overline W}$ and
$\delta {\overline W}^a$ by means of $\delta W$ and $\delta W^a$
(see (4.22) and (4.23) of \cite{xu01})
\begin{eqnarray}
\delta \overline{W} &=& \delta W + \frac{1}{c^2} \left(
 2 \delta W V^2 -4 \delta W^b V^b \right) + O(4) \, , \label{70}\\
\delta \overline{W}^a &=& R ^{ab} \left(
 \delta W ^b - \frac{1}{2} V^b \delta W \right) +O(2) \, .\label{71}
\end{eqnarray}
The inverse transformation takes the form:
\begin{eqnarray}
\delta W &=& \delta \overline{W} + {4 \over c^2} \overline{V}^a
 \delta \overline{W} ^a + O(4) \, , \label{dW} \\
\delta W ^a &=& R^{ba} \left( \delta \overline{W} ^b
 + {1 \over 2}\overline{V}^b \delta \overline{W} \right)
 + O(2)  \, . \label{dWa}
\end{eqnarray}
The transformation formulae is nonsingular, then the behavior of
$\delta \overline{W}$ and $\delta \overline{W}^a$ are similar to
$\delta W$ and $\delta W ^a$, i.e. $\delta \overline{W}$ and
$\delta \overline{W}^a$ are continuous across any interface.
$\delta \overline{W}_{,a}$ and $(\delta \overline{W}^a)_{,b}$ are
finite on the interface, but not necessary continuous. The first
and second time derivatives of $\delta \overline{W}$ and $\delta
\overline{W}^a$ are continuous everywhere including in the
interface since the shock wave does not exist at anytime. Later
when we consider junction condition, we can drop the continuous
terms on both sides of the interface.

For the displacement field $\bf s$ we shall take a similar
physical consideration as in \cite{wahr}. The field $\bf s$ is
continuous across any solid-solid interface. Its normal component
${\bf n} \cdot {\bf s}$ is continuous across any interface. But
its tangent component maybe not continuous for solid-liquid
interface, since the tangent interaction between solid and liquid
is close to zero in the absence of viscous forces. Also $\dot{s}
^a$ and $\ddot{s} ^a$ are finite across any interface. $\mu$,
$\kappa$, $s _{ab}$ and $\Theta $ are finite, but on the different
side of a interface they maybe different (not continuous).
Therefore $\mu _{,a}$, $\kappa _{,a}$, $s_{bc,a}$ and $\Theta
_{,a}$ are not necessary finite on the interface. As we mentioned
before all of layers rotate with the same angular velocity $\bf
\Omega$, so that $V^a$ and $\overline{V}^a ( \overline{V} ^a =
R^{ab}V^b)$ are continuous across the interface. $V^b _{,a}$,
$\overline{V} ^b _{,a}$ and $V^2 _{,a} = \overline{V} ^2 _{,a}$
are continuous as well.

In Eq.(\ref{nab1}) the D'Alembertian can be related to the
non-rotating coordinates. A transformation to rotating coordinates
yields
\begin{equation}\label{nab3}
\nabla ^2 - {1 \over c^2} \frac{\partial ^2}{\partial T^2 } =
  \overline{\nabla} ^2
  + \frac{2\overline{V}^a}{c^2} \frac{\partial^2 }
  {\partial \overline{X}^a \partial \overline{T}}
  - \frac{\overline{V}^a \overline{V}^b}{c^2}
  \frac{\partial^2}{\partial\overline{X}^b\partial\overline{X}^a}
  - {1 \over c^2}\frac{\partial ^2}{\partial \overline{T}^2 }
  \, ,
\end{equation}
where we have used the relation $\partial \overline{V}^a /
\partial \overline{X} ^a = 0$.
Substituting Eq.(\ref{nab3}) and (\ref{dW}) into Eq.(\ref{nab1}),
we get
\begin{eqnarray}\label{dSigma1}
&& \frac{\partial}{\partial \overline{X} ^a} \left[
  \frac{\partial}{\partial \overline{X} ^a} \delta  \overline{W}
  + \frac{2}{c^2}\overline{V} ^a \delta \overline{W}_{,\overline{T}}
  + \frac{4}{c^2} \frac{\partial}{\partial \overline{X} ^a}
  ( \overline{V} ^b \delta \overline{W} ^b )
  - \frac{\overline{V}^a \overline{V}^b}{c^2}
  \frac{\partial\delta\overline{W}}{\partial\overline{X}^b}
  \right] - \frac{1}{c^2}\delta \overline{W} _{,\overline{T}
  \overline{T}} \nonumber \\
&& \hskip12mm = -4\pi G \delta \Sigma +O(4) \, ,
\end{eqnarray}

We also can rewrite $\delta \Sigma $ (to see Eqs.(4.27), (4.31)
and (4.40) of \cite{xu01})
\begin{eqnarray}
\delta \Sigma &=& - \overline{\nabla} \cdot (\rho {\bf s})
  + \frac{1}{c^2} \left[ 3 \rho \overline{V} ^a \dot{s} ^a
  - (p s^a) _{,a} - 3\rho \overline{W} _{,b} s^b
  - \rho \delta \overline{W} \right. \nonumber \\
  && \left. -2\overline{\nabla } \cdot \left( \rho {\bf s}
  (\overline{W} + \overline{V} ^2) \right)
  + \frac{1}{2}\rho \overline{V}^2 {}_{,b} s^b
  - 3\kappa s^b {}_{,b} \right] \, . \label{dSigma2}
\end{eqnarray}
From now on for convenience we omitted ``bar" in $\overline \nabla
$ as we did in \cite{xu01}. Substituting Eq.(\ref{dSigma2}) into
Eq.(\ref{dSigma1}), we get
\begin{equation}\label{ABeq}
\nabla \cdot {\bf A} + B = O(4)
\end{equation}
with
\begin{eqnarray*}
{\bf A} &=&
\nabla \delta \overline{W} - 4 \pi G \rho^* {\bf s} \\
&&  + \frac{1}{c^2} \left[
 2 \overline{\bf V} \delta \overline{W}_{'\overline{T}}
 + 4 \nabla ( \overline{V}^b \delta \overline{W}^b )
 - \overline{\bf V}( \overline{\bf V} \cdot \nabla\delta \overline{W})
 -8\pi G\rho^* {\bf s} ( \overline{W} + \overline{V}^2) \right]
\end{eqnarray*}
and
\[
B = \frac{1}{c^2} \left( 12\pi G\rho {\bf s} \cdot\overline{\bf V}
 -12\pi G \rho {\bf s} \cdot \nabla \overline{W}
 -4\pi G \rho \delta\overline{W}
 + 2\pi G \rho {\bf s} \cdot \nabla\overline{V}^2
 -12\pi G\kappa \Theta
 -\delta \overline{W} _{, \overline{T}\overline{T}}
 \right) \, .
\]
As the discussion on the boundary condition problem in classical
physics, we now integrate Eq.(\ref{ABeq}) over an infinitesimally
small volume $\Delta V$ such that an interface intersects this
volume. We choose $\Delta V$ as a cylinder. The surface (S) of
$\Delta V$ enclose the interface of two different layers with the
shape of a circular drum of a radius $r$ and a depth $h$, where $h
\ll r$ and $r$ is so small that the portion of the interface
contained in (S) can be taken as flat. When $h \rightarrow 0$ the
integration divide into two parts: the first part of
Eq.(\ref{ABeq}) becomes a surface integration by means of Gauss'
theorem, the second part tends to zero with vanishing $\Delta V$
because the terms in $B$ are all continuous or finite. Hence,
\begin{equation}
{\bf n} \cdot {\bf A} |_{\rm layer 1}
  = {\bf n}\cdot{\bf A} | _{\rm layer 2}\, ,
\end{equation}
i.e.
\begin{eqnarray}
&& {\bf n} \cdot \left\{ \nabla \delta \overline{W}
 - 4 \pi G \rho^* {\bf s} + \frac{1}{c^2} \left[
 4 \nabla ( \overline{V}^b \delta \overline{W}_b )
 - \overline{\bf V}( \overline{\bf V} \cdot \nabla\delta \overline{W})
 -8\pi G\rho^* {\bf s} ( \overline{W} + \overline{V}^2)
  \right] \right\} +O(4) \nonumber \\
&& \qquad {\rm continuous \; across \; any \; interface}\, ,
\end{eqnarray}
here we have dropped the term $2\overline{\bf V}
(\delta\overline{W} )_{,T}$ since it is continuous across
interface.

\bigskip
Substituting Eq.(\ref{dWa}) and (\ref{dSiga}) into (\ref{nab2})
and considering $\overline{\nabla}^2=\nabla^2$ and the rotation
matrix $R^{ba}$ dependent on time only, we get
\begin{equation}
\frac{\partial}{\partial \overline{X}^b} \left[
 \frac{\partial}{\partial \overline{X}^b}\left(
  \delta \overline{W} ^a + \frac{1}{2}
  \overline{V}^a \delta \overline{W}\right)
 - 4\pi G s^b \rho \overline{V}^a \right]
 + 4\pi G \left( \rho \dot{s}^a
 + \rho s^b \frac{\partial \overline{V}^a}{\partial \overline{X}^b}
 \right) =O(2) \, .
\end{equation}
Performing a similar integration over an infinitesimal volume as
before, we find that
\begin{equation}\label{vdw}
{\bf n} \cdot \left[ \nabla ( \delta \overline{W}^a
 + \frac{1}{2} \overline{V}^a \delta \overline{W} )
 - 4 \pi G \rho^* {\bf s } \overline{V} ^a \right] +O(2)
 \qquad {\rm continuous \; across \; any \; interface } \, ,
\end{equation}
where $\rho$ has been substituted by $\rho^*$ since this equation
is valid only to $O(2)$. Eq.(\ref{vdw}) can be written by means of
the form of parallel vector, i.e.
\begin{equation}
{\bf n} \cdot \left[ \nabla ( \delta \overline{\bf W}
 + \frac{1}{2} \overline{\bf V} \delta \overline{W} )
 - 4 \pi G \rho^* {\bf s } \overline{\bf V} \right] +O(2)
 \qquad {\rm continuous \; across \; any \; interface } \, .
\end{equation}

Finally a similar integration of Eq.(\ref{euler}) over an
infinitesimal cylindrical volume on the interface leads to
\begin{equation}\label{ABdv}
\int_{\Delta V} (A_{ba,b} + B_a) dV = 0 \, ,
\end{equation}
where
\begin{eqnarray}
A_{ba} &=& -  \kappa \Theta \delta _{ab} - 2 \mu s _{ba}
  + \frac{1}{c^2} \left( 4\mu \overline{W} s_{ba}
  + 2\mu s_{ca} \overline{V}^c \overline{V}^b
  + 4 \kappa \overline{W} \Theta \delta _{ab}
  + \kappa \overline{V} ^d \overline{V} ^c s ^d _{,c} \delta _{ab}
   \right) \, , \label{18b} \\
B_a &=& \rho ^* \left( 1 + \frac{2 \overline{W}_{\rm G}}{
  c^2}\right) \left( \ddot{s} _a
  + 2 \epsilon _{abc} \overline{\Omega}^b \dot{s}^c \right)
  + \rho^* \Theta \overline{W}_{{\rm G},a}
  - \rho ^* s^b _{,a} \overline{W}_{{\rm G},b}
  - \rho ^* ( \delta \overline{W}_{\rm G} ) _{,a}
  - \rho ^* s^b \overline{W}_{{\rm G},b a} \nonumber \\
 &&  + \frac{1}{c^2} \left\{ \rho ^* \left[
   \overline{V} ^a (\overline{V} ^b \ddot{s} ^b )
  + \overline{W}_{{\rm G},b} \dot{s}^b \overline{V} ^a
  - 2 \overline{V} ^b \dot{s}^b \overline{W}_{{\rm G},a}
  + ( \delta \overline{W})_{, \overline{T}}
  \overline{V} ^a + 8 \dot{s}^b \overline{W}_{[b,a]}
  - 4 ( \delta\overline{W} _a ) _{, \overline{T}} \right] \right.
 \nonumber \\
 && \left. - ( 2 \mu s_{ab} )_{,\overline{T} } \overline{V} ^b
 - 2 \mu ( 2 \overline{W}_{,c} s _{ac} + \epsilon _{acb} \Omega ^c
 s_{bd}\overline{V} ^d ) + \kappa ( \Theta \overline{W} _{{\rm G},a}
 - \dot{\Theta} \overline{V}^a )  \right\} \, . \label{18c}
\end{eqnarray}
All of terms in $B_a$ are finite or continuous as we mentioned
before. In terms of median method, we have
\begin{equation}
\int_{\Delta V} B_a dV = \overline{B}_a \pi r^2 h \, .
\end{equation}
The first term of Eq.(\ref{ABdv}) can be deduced as an surface
integration
\begin{equation}
\int_{\Delta V} A_{ba,b} dV = \pi r^2 (n^b_1 A_{ab}
  - n^b_2 A_{ab} )  \, .
\end{equation}
Cancelling $\pi r^2$ and neglecting the higher-order term
$\overline{B}_a h$, we get
\begin{eqnarray}
&& n^b \left\{ \kappa \Theta \delta _{ab} + 2\mu s_{ab}
  - \frac{1}{c^2} \left[ 4\mu \overline{W} s_{ba}
  + 2\mu s_{ca} \overline{V}^c \overline{V}^b
  + 4 \kappa \overline{W}\Theta \delta _{ab}
  + \kappa \overline{V}^d \overline{V}^c s^d{}_{,c}
  \delta _{ab} \right] \right\} +O(4) \nonumber \\
&& \qquad {\rm continuous \; across \; any \; interface} \, .
 \label{Abaeq}
\end{eqnarray}
We should point out that the Newtonian part of Eq.(\ref{Abaeq})
$\kappa \Theta \delta _{ab} + 2\mu s_{ab}$ is just the Newtonian
stress tensor $T_{ab}$ in \cite{wahr}, therefore Eq.(\ref{Abaeq})
is an extended PN version of Wahr's.

\bigskip
The PN junction conditions in Cartesian coordinates are summarized
as follows:
\begin{eqnarray}
&&  {\bf s} \quad {\rm continuous \; across \; any \;
  solid-solid \; interface} \label{19c1} \\
&&  {\bf s} \cdot {\bf n} \quad {\rm continuous \; across \;
  any \; interface } \label{19c2} \\
&& \delta \overline{W} \quad {\rm continuous \; across \; any \;
  interface } \label{19c3} \\
&& \delta \overline{W} ^a \quad {\rm continuous \; across \; any
  \; interface} \label{19c4}  \\
&& {\bf n} \cdot \left\{ \nabla \delta \overline{W}
  - 4\pi G \rho^* {\bf s} + {1 \over c^2} \left[
   4 \nabla \left( \overline{V}^b  \delta \overline{W} _b \right)
  - \overline{\bf V} ( \overline{\bf V} \cdot \nabla
  (\delta\overline{W}) )  - 8\pi G \rho^* {\bf s} ( \overline{W}
  + \overline{V}^2 ) \right] \right\} + O(4) \nonumber \\
&&  \qquad {\rm continuous \; across \;  any \; interface}
 \label{19c5} \\
&& {\bf n} \cdot \left[ \nabla ( \delta \overline{\bf W}
 + \frac{1}{2} \overline{\bf V} \delta \overline{W} )
 - 4 \pi G \rho^* {\bf s } \overline{\bf V} \right] +O(2)
 \qquad {\rm continuous \; across \; any \; interface }
 \label{19c6} \\
&& n^b \left\{ \kappa \Theta \delta _{ab} + 2\mu s_{ab}
 - {1 \over c^2} \left[ 4\mu \overline{W} s_{ab}
 + 2\mu s_{ac} \overline{V}^c \overline{V}^b
 + 4 \kappa \overline{W} \Theta \delta _{ab}
 +  \kappa \delta _{ab} \overline{V}^d \overline{V}^c s ^d {}_{,c}
 \right] \right\}  +O(4) \nonumber \\
 && \qquad {\rm continuous \; across \; any \; interface }
 \label{19c7} \, .
\end{eqnarray}

\bigskip\noindent
For the outer surface boundary conditions we only need simply take
$\mu = \kappa = \rho ={\bf s}= 0 $ outside the elastic body, i.e.
\begin{eqnarray}
&& \delta \overline{W} |_{\rm in}
  = \delta \overline{W} |_{\rm out} \label{surf1}\\
&& \delta \overline{W} ^a|_{\rm in}
  = \delta \overline{W} ^a |_{\rm out} \label{surf2} \\
&& {\bf n} \cdot \left\{ \nabla \delta \overline{W}
  - 4\pi G \rho^* {\bf s} + {1 \over c^2} \left[
   4 \nabla \left( \overline{V}^b  \delta \overline{W}_b \right)
  - \overline{\bf V} ( \overline{\bf V} \cdot \nabla
  (\delta\overline{W}) )  - 8\pi G \rho^* {\bf s} ( \overline{W}
  + \overline{V}^2 ) \right] \right\} |_{\rm in}  \nonumber \\
&& = {\bf n} \cdot \left\{ \nabla \delta \overline{W}
  + {1 \over c^2} \left[
   4 \nabla \left( \overline{V}^b  \delta \overline{W} _b \right)
  - \overline{\bf V} ( \overline{\bf V} \cdot \nabla
  (\delta\overline{W}) ) \right] \right\} |_{\rm out}
  + O(4)\label{surf3} \\
&& {\bf n} \cdot \left\{ \nabla \left( \delta \overline{\bf W}
  + {1 \over 2} \overline{\bf V} \delta \overline{W} \right)
  - 4\pi G \rho^* {\bf s} \overline{\bf V} \right\}|_{in}
 = {\bf n} \cdot \nabla \left( \delta \overline{\bf W}
  + {1 \over 2} \overline{\bf V} \delta \overline{W} \right)|_{out}
  + O(2) \label{surf4}\\
&& n^b \left\{ \kappa \Theta \delta _{ab} + 2\mu s_{ab}
 - {1 \over c^2} \left[ 4\mu \overline{W} s_{ab}
 + 2\mu s_{ac} \overline{V}^c \overline{V}_b
 + 4 \kappa \overline{W} \Theta \delta _{ab}
 +  \kappa \delta _{ab} \overline{V}_d \overline{V}^c s ^d{}_{,c}
 \right] \right\}=O(4) \, . \nonumber \\
&&  \label{surf5}
\end{eqnarray}

\bigskip
For most applications the use of spherical coordinates will be
advantageous. Corresponding boundary and junction conditions cab
be derived similarly to the case of Cartesian coordinates. They
read
\begin{eqnarray}
&&  {\bf D} \quad {\rm continuous \; across \; any \;
  solid-solid \; interface} \label{22a1} \\
&&  {\bf D} \cdot {\bf n} \quad {\rm continuous \; across \;
  any \; interface } \label{22a2} \\
&& \delta \widetilde{W} \quad {\rm continuous \; across \; any
  \; interface } \label{22a3} \\
&& \delta \widetilde{W} ^a \quad {\rm continuous \; across \;
  any \; interface} \; (a=r,\theta, \phi \; {\rm respectively})
  \label{22a4}  \\
&& {\bf n} \cdot \left\{ \nabla \delta \widetilde{W}
  - 4\pi G \rho^* {\bf D} + {1 \over c^2} \left[
   4 \nabla ( \Omega r^2 \sin^2 \theta \delta \widetilde{W}^\phi )
  - 8\pi G \rho^* {\bf D}( \widetilde{W} +
  \Omega ^2 r^2 \sin ^2 \theta ) \right. \right. \nonumber \\
  && \left. \left.  \qquad - {\bf e}_\phi \Omega^2 r \sin{\theta}
  (\delta \widetilde{W} ) _{,\phi} \right] \right\} + O(4)
   \qquad {\rm continuous \; across \;  any \; interface}
 \label{22a5} \\
&& {\bf n} \cdot \left[ \nabla \left( \delta \widetilde{\bf W}
 + {1 \over 2} \widetilde{\bf V} \delta \widetilde{W} \right)
 - 4\pi G \rho^* {\bf D} \widetilde{\bf V} \right] + O(2)
 \qquad {\rm continuous \; across \; any \; interface}
 \label{22a6}\\
&& n^b \left\{ \kappa \Theta D _{ab} + 2\mu s_{ab}
 - {1 \over c^2} \left[ 4\mu \widetilde{W} s_{ab}
 + 2\mu s_{a\phi} \Omega ^2 D_{b \phi}
 + 4 \kappa \widetilde{W} \Theta D _{ab}
 +  \kappa D_{ab} \left(
 \Omega ^2 r \sin \theta D^\phi{}_{,\phi}
 \right. \right. \right. \nonumber \\
 && \left. \left. \left. \qquad
 + \frac{1}{2} {\bf D} \cdot \nabla ( \Omega ^2 r^2 \sin ^2
 \theta ) \right) \right] \right\}
  \qquad {\rm continuous \; across \; any \; interface } \, .
 \label{22a7}
\end{eqnarray}
$s_{ab}$ in Eq.(\ref{22a7}) is the shear tensor in spherical
coordinates, which is shown in the Appendix
Eqs.(\ref{sab1})--(\ref{sab6})

For the outer surface boundary we take $\mu = \kappa = \rho = {\bf
D } = 0$ outside the elastic body as before. When we neglect all
of $1 /c^2$ terms and let $\overline{V}^a =0$(non-rotating), our
formulae (Eq.(\ref{22a5}) and (\ref{22a7})) agree with Wahr's
results (to see Sec. III of Ref.\cite{wahr} ), i.e. our work is an
extension of the Newtonian version to rotating PN version. As for
our formulae (Eq.(\ref{22a6})), since it is a purely PN junction
condition, there is no Newtonian corresponding to be compared.


\section{CONCLUSION}

In this paper we present the perturbation equations for the
dynamical behavior of astronomical elastic bodies in the first
post-Newtonian approximation of Einstein theory of gravity in
rotating spherical coordinates. In comparison with our previous in
Cartesian coordinates\cite{xu01}, these equations in spherical
coordinates are more useful for applications since usually all
relevant equations are expended in terms of scalar, vector and
tensor spherical harmonics. The equations and relations such as
junction conditions can e.g., be applied to problems of
geodynamics (e.g., for the theory of nutation) or seismology of
compact stars (e.g. for the problem of the normal modes of
astronomical bodies). Also the formulation of post-Newtonian
junction conditions at the interface of adjacent layers of
different elastic properties is presented here for the first time.
A comparison reveals that the Newtonian limit agrees with
well-known results from the literature. Here it should be
emphasized that the junction conditions can be written very
generally with ordinary Euclidean 3-vectors and 3-tensors so that
they can be formulated for a broad class of coordinate systems.

Our perturbation equations together with junctions conditions can
in principle be solved if the internal quantities of state
(density, pressure etc.) and elastic moduli are given e.g., by
some Earth's model\cite{ande}. However, these equations describing
the free and forced motions of the body are  complicated partial
differential equations. For that reason usually an expansion of
relevant quantities in terms of spherical harmonics turns these
equations into a set of coupled ordinary differential equations. A
different natural basis for such an expansion is provided by the
so-called generalized spherical harmonics\cite{phin,thor} that was
employed by Wahr and other authors. By using the generalized
spherical harmonics
 to expand all functions (displacement vector,
incremental Eulerian gravitational potential energy, incremental
elastic stress tensor, applied force (tidal force et al.)), the
original partial differential equations and boundary conditions
are transformed into a set of ordinary differential equations and
scalar boundary conditions for the unknowns functions. For a
spherical and non-rotating ground state these ordinary
differential equations are uncoupled, for an oblate, rotating body
they are coupled, the coupling parameter being given by the
dimensionless oblateness of the body. Such expansions together
with applications for geodynamics will be published separately.


\acknowledgments

This work was supported by the National Natural Science Foundation
of China (Grant No. 10273008) and the German Science Foundation
(DFG).

\appendix*
\section{Explicit results for the strain and shear tensor}

\bigskip\noindent
From (\ref{emunu}) one derives
\begin{eqnarray}
e _{00} &=& O(4) \, ,  \\
e _{0r} &=& e _{0\theta} = e_{0 \phi} = O(5)  \, , \\
e _{rr} &=& D ^r_{,r} + {1 \over c^2} \left( \delta \widetilde{W}
 + {\bf D} \cdot \nabla \widetilde{W}
  + 2\widetilde{W} D^r_{,r} \right) + O(4) \, , \\
e _{\theta\theta}
 &=& r^2 \left\{ {1 \over r} D ^\theta _{,\theta}
 + {1 \over r} D^r + {1 \over c^2} \left( \delta{\widetilde{W}}
 + {2\widetilde{W} \over r} D^r + {\bf D} \cdot \nabla \widetilde{W}
 + {{2\widetilde{W} } \over r} D^\theta _{,\theta} \right) \right\}
 + O(4) \, ,  \\
e _{\phi\phi}
 &=& r^2 \sin^2\theta \left\{ \frac{1}{r\sin\theta} D ^\phi _{,\phi}
 + {1 \over r} D^r + {1 \over r}D^\theta \cot\theta
 + {1 \over c^2} \left[ \delta{\widetilde{W}}
 + {\bf D} \cdot \nabla \widetilde{W}
 \right. \right. \nonumber \\
 && \hskip-4mm \left. \left.
 + \frac{1}{r \sin\theta} (2\widetilde{W} + V ^2 ) D^\phi _{,\phi}
 + \Omega r \sin\theta D ^\phi _{,T}
 + {2 \over r} ( \widetilde{W} + V ^2 ) (D^r + D^\theta \cot\theta )
 \right] \right\} + O(4) ,  \\
e _{r\theta} &=& r \left\{ {1 \over 2r} ( D^r _{,\theta}
 + r D^\theta _{,r} - D^\theta ) \left( 1 + \frac{2\widetilde{W} }{c^2}
 \right) \right\} + O(4) \, , \\
e _{r \phi} &=& r \sin\theta \left\{
 {1 \over 2} \left( \frac{D^r_{,\phi}}{r \sin\theta}
 - {1 \over r} D^\phi + D^\phi _{,r} \right)
 + {1 \over c^2} \left[ \frac{\widetilde{W} }{r \sin\theta} D^r_{,\phi}
 \right. \right. \nonumber \\
 && \hskip-3mm \left. \left.
 + ( D^\phi _{,r} - \frac{D^\phi}{r})(\widetilde{W} + {1 \over 2} V^2 )
 + {1 \over 2} \Omega r \sin\theta D^r_{,T} \right]
 \right\} + O(4) \, , \\
e _{\theta\phi} &=& r^2 \sin\theta \left\{
 {1 \over 2} \left( \frac{D^\theta _{,\phi}}{r \sin\theta}
 - \frac{\cot\theta}{r} D^\phi + {1 \over r} D^\phi _{,\theta}
 \right) + {1 \over c^2}
 \left[ \frac{\widetilde{W} }{r \sin\theta} D^\theta _{,\phi}
 \right. \right. \nonumber \\
 && \hskip-3mm \left. \left.
 + \left( \widetilde{W} + {1 \over 2} V^2 \right)
 \left( \frac{D^\phi _{,\theta}}{r}
 - \frac{\cot\theta}{r} D^\phi \right)
 + {1 \over 2} \Omega r \sin\theta D^\theta _{,T} \right]
 \right\} + O(4)
\end{eqnarray}
where
\[ {\bf D} \cdot \nabla \widetilde{W} =
  \widetilde{W} _{,r} D^r + {1 \over r} \widetilde{W} _{,\theta} D ^\theta
 + \frac{1}{r \sin\theta} \widetilde{W} _{,\phi} D^\phi \, . \]

\medskip\noindent
The volume dilatation $\Theta$, which is related with the
expansion rate $\theta $ by $\theta = \Theta _{,\mu} U ^\mu$,
reads
\begin{eqnarray}
\Theta &=& e ^\mu{}_\mu = \widetilde{G} ^{\mu\nu} e_{\mu\nu}
 = \widetilde{G} ^{ab} e _{ab} + O(4) \nonumber \\
 &=& \nabla\cdot {\bf D} + {1 \over c^2} \left(
 3\delta{\widetilde{W}} + 3 {\bf D} \cdot \nabla \widetilde{W}
 + \Omega r \sin\theta D^\phi _{,T}
 + {1 \over 2} V ^2_{,c} \xi ^c  \right) + O(4) \, ,
\end{eqnarray}
where
\begin{eqnarray*}
\nabla \cdot {\bf D} & \equiv & D^r_{,r}
 + {1 \over r} D^\theta _{,\theta}
 + \frac{1}{r\sin\theta} D^\phi _{,\phi}
 + {2 \over r} D^r + {1 \over r} \cot\theta D^\theta \, ,\\
{1 \over 2} V ^2_{,c} \xi ^c &=& \Omega ^2 r\sin^2\theta \xi^r
 + \Omega ^2 r^2 \sin\theta\cos\theta \xi ^\theta
 = \Omega ^2 r\sin^2\theta (D^r + \cot\theta D ^\theta ) \, .
\end{eqnarray*}
From this it is easy to derive the symmetric trace-free shear
tensor $s_{\mu\nu}$. One finds
\begin{eqnarray}
s_{00} &=& O(4) \, ,  \label{sab1} \\
s_{0r} &=& s_{0\theta} = s_{0\phi} = O(5) \, , \label{sab2} \\
s_{rr} &=& D ^r_{,r} - {1 \over 3} \nabla \cdot {\bf D}
 + {1 \over c^2} \left\{ 2 \widetilde{W} \left( D^r _{,r}- {1 \over 3} \Theta
 \right) \right. \nonumber \\
 && \hskip-3mm \left. - {1 \over 3} \left( \Omega r \sin\theta D^\phi _{,T}
 + \Omega ^2 r\sin^2\theta (D^r + \cot\theta D ^\theta )\right)
 \right\} + O(4) \, , \label{sab3} \\
s_{\theta\theta}
 &=& r^2 \left\{ {1 \over r} D ^\theta _{,\theta}
 + {1 \over r} D^r - {1 \over 3} \nabla \cdot {\bf D}
 + {1 \over c^2} \left[ 2\widetilde{W} \left( {1 \over r} D ^\theta _{,\theta}
 + {1 \over r} D^r - {1 \over 3} \Theta \right) \right. \right.
 \nonumber \\
 && \hskip-3mm \left. \left.
 - {1 \over 3} \left( \Omega r \sin\theta D^\phi _{,T}
 + \Omega ^2 r\sin^2\theta (D^r + \cot\theta D ^\theta ) \right)
  \right] \right\} + O(4) \, , \label{sab4} \\
s_{\phi\phi}
  &=& r^2 \sin^2\theta \left\{ \frac{1}{r\sin\theta} D ^\phi _{,\phi}
 + {1 \over r} D^r  + {1 \over r}D^\theta \cot\theta
 - {1 \over 3} \nabla \cdot {\bf D} \right. \nonumber \\
 && \hskip-3mm \left. + {1 \over c^2} \left[ ( 2\widetilde{W} + V ^2 )
 \left( \frac{1}{r\sin\theta} D ^\phi _{,\phi}
 + {1 \over r} D^r  + {1 \over r}D^\theta \cot\theta
 - {1 \over 3} \Theta \right) \right. \right. \nonumber \\
 && \hskip-3mm \left. \left. + \Omega ^2 r \sin^2\theta \left( D^r
 + \cot\theta D^\theta \right) + {2 \over 3} \Omega r \sin\theta
 D^\phi _{,T}
 \right. \right. \nonumber \\
 && \hskip-3mm \left. \left. - {1 \over 3} \Omega ^2 r\sin^2\theta
 (D^r + \cot\theta D ^\theta ) \right] \right\}
 + O(4) \, , \label{sab5}
\end{eqnarray}
and
\begin{equation}\label{sab6}
s_{r\theta} = e_{r \theta} \, , \qquad s_{r\phi} = e_{r \phi} \, ,
\qquad s_{\theta\phi} = e_{\theta\phi} \, .
\end{equation}
These are
the post-Newtonian components of the Lagrangian strain  and
shear tensor in  spherical coordinates. If all post-Newtonian
terms are neglected (i.e., for  $c \rightarrow \infty $), they
reduce to the well known Newtonian results that  can be found in
the standard literature.




\begin{thebibliography}{}

\bibitem
[1] {jeff} H.~Jeffreys and R.O.~Vicente, Mon. Not. R. Astro. Soc.
{\bf 117}, 142 (1957).

\bibitem
[2] {mori} H.~Moritz and I.I.~Mueller, {\it Earth Rotation-Theory
and Observation} , Ungar Pub. Co., New York, NY10017, (1987).

\bibitem
[3] {wahr} J.M.~Wahr, Geodesy and Geo-dynamics (H.~Moritz and
H.~S\"ukel (eds), Mitteilungen der Geod\"atischen Institute der
Technischen Universit\"at graz, No.41), pp.327-379 (1982).

\bibitem
[4] {scha} J.~Schastok, Geophys. J. Int., {\bf 130}, 137 (1997).

\bibitem
[5] {deha} V.~Dehant and P.~Defraigne, J. Geophys. Res., {\bf 102
(B12)}, 27, 659, 687 (1997).


\bibitem
[6] {hugh} S.A.~Hughes and K.S.~Thorne, Phys. Rev. D{\bf 58},
122002 (1998).

\bibitem
[7] {chin} E.S.C.~Ching et al.,
 preprint, 1 - 14 (2002).
\bibitem
[8] {card} V.~Cardoso and J.P.S.~Lemos, Clas. Quant. Grav. {\bf
18}, 5257 (2001).

\bibitem
[9] {xu97} C.~Xu, X.~Wu and G.~Sch\"afer, Phys. Rev. D{\bf 55},
528 (1997).

\bibitem
[10] {xu01} C.~Xu, X.~Wu and M.~Soffel,
 Phys. Rev. D{\bf 63}, 043002 (2001).

\bibitem
[11] {damo91} T.~Damour, M.~Soffel and C.~Xu, Phys.~Rev.~D {\bf
43}, 3273 (1991).

\bibitem
[12] {damo92} T.~Damour, M.~Soffel and C.~Xu, Phys.~Rev.~D {\bf
45}, 1017 (1992).

\bibitem
[13] {damo93} T.~Damour, M.~Soffel and C.~Xu, Phys.~Rev.~D {\bf
47}, 3124 (1993).

\bibitem
[14] {damo94} T.~Damour, M.~Soffel and C.~Xu, Phys.~Rev. D {\bf
49}, 618  (1994).

\bibitem
[15] {cart72} B.~Carter and H.~Quintana, Proc. Roy. Soc. Lond.
{\bf A331}, 57 (1972).

\bibitem
[16] {cart73} B.~Carter, Commun. Math. Phys. {\bf 30}, 261 (1973).

\bibitem
[17] {lapw} E.R.~Lapwood and T.~Usami, Free Oscillation of the
Earth, Cambridge Univ. Press. Chapt. 7 (1981).

\bibitem
[18 ] {lich} A.~Lichnerowicz, Theories relativistes de la
gravitation at de electromagnetisme. Masson, Paris,
 Chap. I, III (1955).

\bibitem
[19] {robs} E.H.~Robson, Ann. Inst. Henri Poincare', {\bf 15}, 41
 (1972).

\bibitem
[20] {ande} D.L.~Anderson, Phys. Earth Planet Inter., {\bf 25},
297 (1981).

\bibitem
[21] {phin} R.A.~Phinney and R.~Burridge, Geophys. J. R. Astro.
Soc, {\bf 34}, 451 (1973).

\bibitem
[22] {thor} K.S.~Thorne, Rev. Mod. Phys, {\bf 52}, 299 (1980).

\end{thebibliography}
\end{document}